\documentclass[review,12pt]{elsarticle}
\usepackage{lineno,hyperref}
\modulolinenumbers[5]
\journal{Journal of elsevier  \LaTeX\ Templates}
\usepackage{amsmath}
\usepackage{graphicx}

\title{Amplification of Acoustic Waves in Armchair Graphene Nanoribbon in the 
Presence of External Electric and Magnetic Field}

\author[rvt]{K. A. Dompreh\corref{cor1}\fnref{fn1}}
\ead{kwadwo.dompreh@ucc.edu.gh}
\author[rvt]{S. Y. Mensah}
\author[rvt]{S. S. Abukari}
\author[rvt]{F. Sam}
\author[focal]{N. G. Mensah}
\address[rvt]{Department of Physics, College of Agriculture and Natural Sciences, U.C.C, Ghana.}
\address[focal]{Department of Mathematics, College of Agriculture and Natural Sciences, U.C.C, Ghana}
\cortext[cor1]{Corresponding author\\ Email: kwadwo.dompreh@ucc.edu.gh}
\ead[url]{kwadwo.dompreh@ucc.edu.gh}

\date{}
\begin{document}
\begin{abstract}
Amplification of Acoustic Waves in Armchair Graphene Nanoribbon (AGNR) in the presence of an external Electric and Magnetic field 
was studied using the Boltzmann's kinetic equation. The general 
expression  for the Amplification $(\Gamma_{\perp}/\Gamma_0)$ was obtained in the region $ql >> 1$ for the energy dispersion $\varepsilon(\vec{p})$
near the Fermi point. For various parameters of the quantized wave vector ($\beta$), the analysis of
$\Gamma_{\perp}/\Gamma_0$ against the sub-bands index $(p_i)$; width of AGNR; and magnetic strength $(\Omega\tau)$, were numerically analyzed. The results showed a 
linear relation for $\Gamma_{\perp}/\Gamma_0$ with constant electric field $(\vec{E})$ but non-linear
for $\Gamma_{\perp}/\Gamma_0$  with $q$ or $\Omega\tau$. Sound Amplification  in AGNR is reported with an increase in 
Acoustic wave number $(\vec{q})  > 1.5\times 10^{7}cm^{-1}$. This can cause SASER in 
Armchair Graphene Nanoribbon (AGNR).
\end{abstract}

\maketitle

\section*{Introduction}
In Semiconductors materials, it is well known that the interaction of  phonons  with 
conducting electrons can lead to either absorption or amplification of  phonons~\cite{1,2}. 
The absorption  of momentum from the phonons  to the charge carriers  results in the 
generation of acoustic effect such as (I) Acoustoelectric effect~\cite{3,4,5,6,7,8,9,10} 
(II) Acoustomagnetoelectric effect~\cite{11,12,13,14,15}
and (III) Acoustothermal effect~\cite{16}. Amplification on the other hand occur when the charge
carriers loses energy to the phonons. This phenomena has been extensively studied  both theoretically and 
experimentally in bulk semiconductors~\cite{17,18,19}. Tolpygo (1956), Uritskii~\cite{20},
and  Weinreich~\cite{21} theoretically studied acoustic wave amplification in semiconductors and was experimentally
observed in CdS by Hudson  ~\cite{22}  and in N-Ge by Pomeranztz~\cite{23}. 
In low-dimensional systems, the acoustic wave amplification (absorption) was studied theoretically  and 
experimentally~\cite{24,25,26,27,28}. 
Recently the study of acoustic effect in semiconductor nanostructure materials is extended to Carbon Nanotube (CNT)~\cite{29,30,31,32} 
and Graphene with few experimental work carried out~\cite{33,34,35,36}. These carbon based materials have interesting properties as well   
as  an excellent combination of electronic, 
optoelectronic, and thermal properties compared to conventional rigid silicon which 
makes them excellent systems for application in electronic and optoelectronic systems. 
Graphene~\cite{37} is a single-atom sheet of graphite.  The most interesting property of Graphene is its  linear 
energy dispersion  $E = \pm \hbar V_F \vert k \vert$ (the Fermi velocity $V_F \approx 10^8ms^{-1}$) at the Fermi level with 
low-energy excitation. Graphene-based electronics has attracted much attention due to high carrier mobility in
bulk graphene devices such as sub-terahertz  Field-effect transistors~\cite{38}, infrared transparent electrodes~\cite{39} 
and $THz$ plasmonic devices~\cite{40}. 
However, charge transport in Graphene differ from other $2$D systems due to lack of an electronic band gap which makes graphenes 
unsuitable for use as semiconducting electronic devices~\cite{41}. To overcome this problem, strips of Graphene called Graphene Nanoribbons (GNRs) 
whose characteristics are dominated by the nature of their edges are proposed \cite{42,43}. The Armchair Graphene Nanoribbon (AGNR) and 
Zigzag Graphene Nanoribbon (ZGNR) with well-defined width have being extensively studied using the tight-binding approach \cite{44} and 
Edge-functionalization method in Density Functional Theory (DFT)\cite{45}. The -H, -F, -Cl, -Br, -S, -SH and -OH are used to engineer 
the band gap so as to utilised the various properties of GNRs for electronic applications.

In this paper, the Boltzmann kinetic equation for electron-phonon interactions is used to study the Acoustic Wave Amplification in AGNR.
This is achieved by applying a sound flux to the AGNR in the presence of external d.c. electric and a constant magnetic field.
In this work, it is noted that increasing the acoustic wave number $(\vec{q})$, causes 
intraband transition leading the attainment of phonon application in AGNR. 
Varying other parameters such as the sub-band index $(p_i)$, the width of AGNR, the energy gap $(E_g)$ and 
the magnetic strength $(\Omega\tau)$ also modulate the Amplification.  This paper is organised as follows: section $1$ deals with 
the introduction;  section $2$, the theory of the acoustic wave amplification; section $3$ is the numerical results and   discussions;
 whilst section $4$ is the conclusion.  

\section*{Theory}
To calculate for the amplification of acoustic waves in AGNR, we shall proceed from  the method developed in \cite{46}, where the sound flux $(\vec{W})$, 
d.c. electric field $(\vec{j})$ and a constant magnetic field $(\vec{H})$ are considered mutually perpendicular to the plane of the AGNR. 
The acoustic wave is considered in the hypersound regime $ql >> 1$ ($q$ is the acoustic wavenumber and $l$ is the mean free path of an electron). 
To solve for the partial current generated in the AGNR, the Boltzmann kinetic equation 
\begin{eqnarray}
-\left(e\vec{E}\frac{\partial f_{\vec{p}}}{\partial{\vec{p}}} + \Omega[\vec{p},\vec{H}],\frac{\partial f_{\vec{p}}}{\partial {\vec{p}}}\right)=
 -\frac{f_{\vec{p}} - f_0(\varepsilon_{\vec{p}})}{\tau(\varepsilon_{\vec{p}})}  +\nonumber\\
  \frac{\pi {\xi}^2 \vec{W}}{\rho V_s^3} \left\{ {[f_{{\vec{p}}+{\vec{q}}}} - f_{\vec{p}}]\delta(\varepsilon_{\vec{p}+\vec{q}} 
-\varepsilon_{\vec{p}}- \hbar\omega_{\vec{q}})+ {[f_{{\vec{p}}-{\vec{q}}}} - f_{\vec{p}}]\delta(\varepsilon_{\vec{p}-\vec{q}} 
-\varepsilon_{\vec{p}}+\hbar\omega_{\vec{q}})\right\} \label{EQ_1} 
\end{eqnarray}
is employed. Here, $\xi$ is the constant of deformation potential, $e$ the electronic charge, $\vec{E}$ is the constant electric field produced by the acoustic wave in the open-circuited field, $\omega_{\vec{q}}$
is frequency, $\vec{W}$ is the density of the acoustic flux, $\rho$ is the density of the sample and $\vec{p}$ the characteristic quasi-momentum of the electron. 
The relaxation time on energy is $\tau(\varepsilon_{\vec{p}})$ and the cyclotron frequency, $\Omega = \frac{\mu H}{c}$. where $\mu$ is the mobility,
$c$ is the speed of light, $\bar h$ is the planck constant,  $f_0(\varepsilon)$  is the equilibrium function of the electron distribution, and $\vec{q}$ is the acoustic wavenumber of the sound.
The energy dispersion relation $\varepsilon(\vec{p})$
for AGNRs band near the Fermi point is expressed \cite{47} as 
\begin{equation}
 \varepsilon(\vec{p}) = \frac{E_g}{2} \sqrt{[(1+\frac{\vec{p}^2}{{\hbar}^2{\beta}^2})]} \label{EQ_2}
\end{equation}
where the energy gap $E_g =3ta_{c-c}\beta$, with $\beta$ being the quantized wave vector given as
\begin{equation}
\beta =\frac{2\pi}{a_{c-c}\sqrt{3}}\left(\frac{p_i}{N+1}- \frac{2}{3}\right)\label{EQ_3}
\end{equation}
The $p_i$ is the subband index, $N$ the number of dimmer lines which determine the width of the AGNR,
$a_{c-c} = 1.42\dot{A}$ is the Carbon-Carbon (C-C) bond length, $t = 2.7 eV$ is the nearest neighbor 
(C-C) tight binding overlap energy. The distribution function $f_p(\varepsilon)$ is expressed by Taylor expansion as 
\begin{equation}
f_{\vec{p}} = f_0(\varepsilon)- \vec{p}f_1(\varepsilon)+...\label{EQ_4}
\end{equation}
The ${f_1}(\varepsilon) = \vec{\chi}(\varepsilon)\frac{\partial f_0}{\partial \varepsilon}$  is the perturbative part.
The $\vec{\chi}(\varepsilon)$ characterises the deviation of the  $f_p$ from its equilibruim 
and is determined from the Boltzmann kinetic equation. 
Multiplying the Eqn.(\ref{EQ_1}) by $\vec{p}\delta(\varepsilon - \varepsilon_{\vec{p}})$ and summing over
$\vec{p}$ reduces the Boltzmann kinetic equation to
\begin{equation}
 \frac{{\vec{R}(\varepsilon)}}{\tau(\varepsilon)} + \Omega\left[{\vec{H}},{\vec{R}(\varepsilon)}\right] 
= {\vec{\Lambda}(\varepsilon)}+{\vec{S}(\varepsilon)}  \label{EQ_5}
 \end{equation}
 where $\vec{R}(\varepsilon)$ is the partial current density given as 
\begin{equation}
{\vec{R}}(\varepsilon) \equiv e\sum_{\vec{p}} \vec{p} f_{\vec{p}} \delta(\varepsilon -\varepsilon_{\vec{p}}) \label{EQ_6}
\end{equation}
with $\vec{\Lambda}(\varepsilon)$ and $\vec{S}(\varepsilon)$ given as 
\begin{equation}
{\vec{\Lambda}}(\varepsilon) =-e\sum_{\vec{p}}\left({\vec{E}},\frac{\partial f_{\vec{p}}}{\partial{\vec{p}}}\right)
{\vec{p}}\delta(\varepsilon -\varepsilon_{\vec{p}}) \label{EQ_7}
\end{equation}
\begin{eqnarray}
{\vec{S}}{(\varepsilon)} = \frac{\pi {\xi}^2 \vec{W}}{\rho V_s^3}\sum_{\vec{p}}\vec{p}\delta(\varepsilon -\varepsilon_{\vec{p}})\{ {[f_{{\vec{p}}+{\vec{q}}}} - f_{\vec{p}}]\delta(\varepsilon_{\vec{p}+\vec{q}} -\varepsilon_{\vec{p}}-\hbar\omega_{\vec{q}})
 +{[f_{{\vec{p}}-{\vec{q}}}} - f_{\vec{p}}]\nonumber\\
\delta(\varepsilon_{\vec{p}-\vec{q}} -\varepsilon_{\vec{p}}+\hbar\omega_{\vec{q}})\}  \label{EQ_8}
\end{eqnarray}
Considering ${\vec{p} \rightarrow -{\vec{p}}}$,  $f_{\vec{p}} \rightarrow  f_0(\varepsilon_{\vec{p}})$, 
by transforming the summation into integrals and integrating gives 
\begin{eqnarray}
\vec{\Lambda}(\varepsilon) =\vec{E}\left(\frac{2{\hbar}^2\beta^2}{{\hbar}\vec{q}}\alpha - \frac{{\hbar}\vec{q}}{2}\right)
\frac{\partial {f_0}}{\partial{\varepsilon}}\frac{\Theta \left(1 -{\alpha}^2\right)}{\sqrt{1- {\alpha}^2}}\label{EQ_9}
\end{eqnarray}
\begin{eqnarray}
{\vec{S}}({\varepsilon}) = \frac{2\pi{{\vec{W}}}\phi}{\rho V_s\alpha}\left(\frac{2\hbar^2\beta^2}{\hbar q}\alpha -\frac{\hbar q }{2}\right)
 \frac{\Theta \left(1 -{\alpha}^2\right)}{\sqrt{1- {\alpha}^2}}\frac{1}{f_0(\varepsilon)}\frac{\partial f_0}{\partial {\varepsilon}} \label{EQ_10}
\end{eqnarray}
where $\alpha = \frac{\hbar\omega_q}{E_g}$,  $\phi =\frac{E_g^2 \alpha^2}{2 V_s^2}f_0(\varepsilon)$ and $\Theta$ is the Heaviside step function represented as
\begin{equation*}
\Theta(1-\alpha^2)=
\begin{cases}{
1 \quad \text{ if $(1-\alpha^2) > 0$}}\\
{0 \quad \text{if $(1-\alpha^2) < 0$}
}\end{cases}
\end{equation*}
Substituting Eqn.(\ref{EQ_9}) and Eqn.(\ref{EQ_10}) into Eqn.(\ref{EQ_5}) and solving for $\vec{R}(\varepsilon)$ gives
\begin{eqnarray} 
\vec{R}(\varepsilon) =\{\frac{2\pi\phi}{\rho
V_s\alpha}\left(\frac{2\hbar^2\beta^2}{\hbar q}\alpha -\frac{\hbar q }{2}\right) 
\frac{1}{f_0(\varepsilon)}\frac{\partial f_0}{\partial {\varepsilon}}\nonumber\\
\{\vec{W}\tau(\varepsilon)+\vec{\Omega}[\vec{h},\vec{W}]\tau^2(\varepsilon)+  \Omega^2\vec{h}(\vec{h},\vec{W})\tau^3(\varepsilon)\}+\left(\frac{2\hbar^2\beta^2}{\hbar q}\alpha -\frac{\hbar q }{2}\right)\frac{\partial f_0}{\partial {\varepsilon}}\nonumber\\
\{\vec{E}\tau(\varepsilon)+\Omega[\vec{h},\vec{E}]\tau^2(\varepsilon) +  \Omega^2 \tau^3 \vec{h}(\vec{h},\vec{E})\}\}\frac{\Theta \left(1 -{\alpha}^2\right)}{\sqrt{1- {\alpha}^2}}\{1+\Omega^2\tau^2(\varepsilon)\}^{-1}\label{EQ_11}
\end{eqnarray} 
which can be written as  $\vec{R}(\varepsilon) = \vec{\chi}(\varepsilon)\frac{\partial f_0}{\partial {\varepsilon}}$. 
In the linear approximation of $\vec{E}$, the acoustic flux $\vec{W}$ and the term $\Omega^2  \vec{h}(\vec{h},\vec{E})\tau^3(\varepsilon)$ are ignored. This makes 
$\vec{\chi}(\varepsilon)$ reduces to
\begin{eqnarray}
\vec{\chi}(\varepsilon) = \{\vec{E}\tau(\varepsilon)+\Omega[\vec{h},\vec{E}]\tau^2(\varepsilon)+
\Omega^2  \}\left(\frac{2\hbar^2\beta^2}{\hbar q}\alpha -\frac{\hbar q }{2}\right)\{1+\Omega^2\tau^2(\varepsilon)\}^{-1} \label{EQ_12}
\end{eqnarray}
For arbitrary orientation of fields,  the current density $\vec{j}$ is 
\begin{equation} 
\vec{j}=-\int_0^{\infty}{\vec{R}(\varepsilon)d\varepsilon} \label{EQ_13}
\end{equation}
Substituting Eqn.(\ref{EQ_12}) and averaging over energy gives 
\begin{equation}
\vec{j}=\{\langle {\frac{\tau(\varepsilon)}{1+\Omega\tau^2(\varepsilon)}}\rangle\vec{E}_y - \langle\frac{\tau^2(\varepsilon)}{1+\Omega\tau^2(\varepsilon)}\rangle [\Omega,\vec{E}]_y\}
\left(\frac{2\hbar^2\beta^2}{\hbar q}\alpha -\frac{\hbar q }{2}\right)\{1+\Omega^2\tau^2(\varepsilon)\}^{-1}  \label{EQ_14}
\end{equation}
Solving for $\vec{E_y}$ in Eqn.(\ref{EQ_14}) in an open circuit system  $(\vec{j}_y = 0)$, 
and substituting into Eqn.(\ref{EQ_12}) for $\langle\langle\vec{\chi}(\varepsilon)\rangle\rangle_y$  yields
\begin{equation}
\langle\langle\vec{\chi}(\varepsilon)\rangle\rangle_y = \Omega \vec{E_x}[\langle\langle\frac{\tau(\varepsilon)}{1+\Omega^2\tau^2(\varepsilon)}\rangle\rangle \frac{\langle\frac{\tau^2(\varepsilon)}{1+\Omega^2\tau^2(\varepsilon)}\rangle}
{\langle\frac{\tau(\varepsilon)}{1+\Omega^2\tau^2(\varepsilon)}\rangle}- \langle\langle\frac{\tau^2(\varepsilon)}{1+\Omega^2\tau^2(\varepsilon)}\rangle\rangle] \label{EQ_15}
\end{equation}
In Eqn.(\ref{EQ_15}) the following averages were used 
\begin{displaymath}
\langle\langle\frac{\tau^k(\varepsilon)}{1+\Omega\tau^2(\varepsilon)}\rangle\rangle
  =-\frac{2\pi}{f_0({\varepsilon})}{\int_{0}^{\infty}} (\frac{\tau^k(\varepsilon)}{1+\Omega\tau^2(\varepsilon)})\frac{\partial f_0}{\partial{\varepsilon}}d{\varepsilon}
\end{displaymath}
\begin{displaymath}
 \langle\frac{\tau^k(\varepsilon)}{1+\Omega\tau^2(\varepsilon)}\rangle
 =-{\int_{0}^{\infty}} (\frac{\tau^k(\varepsilon)}{1+\Omega\tau^2(\varepsilon)})\frac{\partial f_0}{\partial{\varepsilon}}d{\varepsilon}
\end{displaymath}
where $k = 1, 2, 3$ and  $f_0 =[1-exp(\frac{-1}{k_B T}(\varepsilon - \varepsilon _F))]^{-1}$ is the 
Fermi-Dirac distribution function. $\varepsilon _F$ is the Fermi energy, $k_{\beta}$ the Boltzmann constant and $T$ the absolute temperature.

\section*{Sound Absorption}
The general formula for the electronic sound absorption coefficient $(\Gamma_{\vec{q}})$ has the form
\begin{equation}
\Gamma(\vec{q}) = \Gamma_0[1 -\frac{1}{V_s \vec{q} f_0(\varepsilon)}\int_{\varepsilon}^{\infty} \vec{q}f_1(\varepsilon_p)d{\varepsilon}] \label{EQ_16}  
\end{equation}
where $\Gamma_0$ is the absorption coefficient in the absence of external fields, $V_s$  is the speed of sound. 
From Eqn.(\ref{EQ_4}), the above  Eqn.(\ref{EQ_16}) is equivalent to 
\begin{equation}
\Gamma({q}) = \Gamma_0[1 -\frac{(q, \langle\langle\chi(\varepsilon)\rangle\rangle)}{qV_s}]  \label{EQ_17} 
\end{equation}
Inserting Eqn.(\ref{EQ_15}) into Eqn.(\ref{EQ_17}) gives the sound amplification $(\Gamma_{\perp})$ perpendicular to the electric current 
$(\vec{j})$  as 
\begin{equation}
\Gamma_{\perp}=\Gamma_0\{1 -\frac{\Omega \vec{E_x}}{V_s}[\langle\langle\frac{\tau(\varepsilon)}{1+\Omega^2\tau^2(\varepsilon)}\rangle\rangle \frac{\langle\frac{\tau^2(\varepsilon)}{1+\Omega^2\tau^2(\varepsilon)}\rangle}
{\langle\frac{\tau(\varepsilon)}{1+\Omega^2\tau^2(\varepsilon)}\rangle}- \langle\langle\frac{\tau^2(\varepsilon)}{1+\Omega^2\tau^2(\varepsilon)}\rangle\rangle]\} \label{EQ_18}
\end{equation}  
\section*{Numerical analysis}
In solving for Eqn.({\ref{EQ_18}), the following were assumed: At low temperature $kT << 1$, and $\frac{\partial f_0}{\partial \varepsilon} = 
\frac{-1}{k_{\beta}T}exp(-\frac{\varepsilon-\mu}{k_{\beta} T})$.
The final equation therefore simplifies to
\begin{eqnarray}
\Gamma_{\perp}/\Gamma_0 = [1-\frac{9\pi^2}{8V_s}\Omega E_x \tau^2 exp(\Omega^2)\{\frac{3\pi}{16}\frac{F_{(-3/2,\Omega^2)}F_{(-1/2,\Omega^2)}}{F_{(-2,\Omega^2)}}-F_{(0,\Omega^2)}\}\nonumber\\
\left(\frac{{2\hbar}^2\beta^2}{\hbar q}\alpha-\frac{\hbar q }{2}\right)] \label{EQ_19}
\end{eqnarray}
where $F_{m,n} = \int_0^{\infty}{\frac{x^m}{1+\Omega^2 x^n}}\frac{\partial f_0(\varepsilon)}{\partial x}dx$.
\begin{figure}
\includegraphics[width=6.5cm]{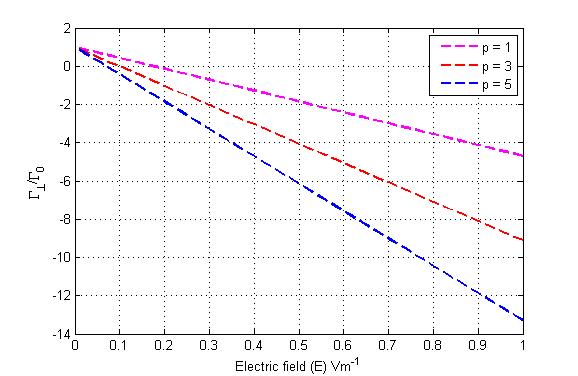}
\includegraphics[width=6.5cm]{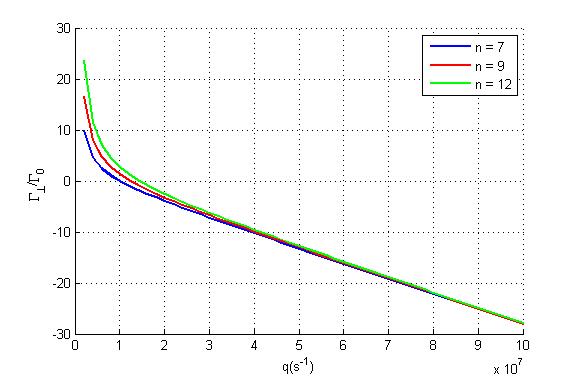}
\caption{(a) Dependence of $\Gamma/\Gamma_0$ on $E_0$ for $7$-AGNR ($p = 2, 4, 6, 7$). 
(b) Dependence of $\Gamma/\Gamma_0$ $q$ at widths of AGNR = $7, 9, 12$.} 
\end{figure}
\begin{figure}
\includegraphics[width=6.5cm]{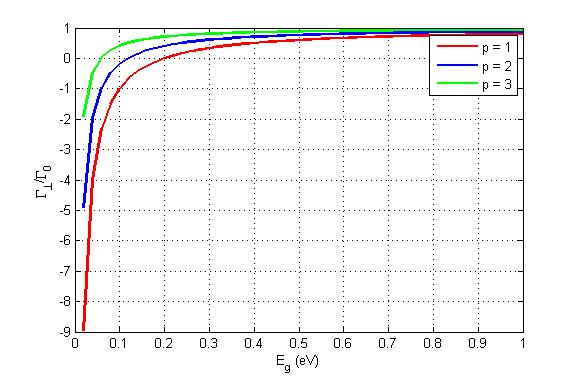}
\includegraphics[width=6.5cm]{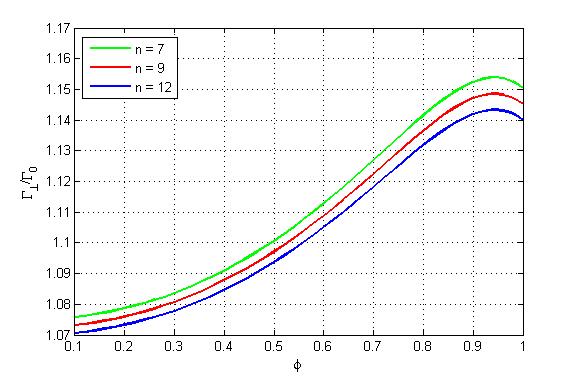}
\caption{(a) Dependence of $\Gamma/\Gamma_0$ on the $7$-AGNR energy gap ($E_g$) at $p_i = 1, 2, 3$  
(b) Dependence of $\Gamma/\Gamma_0$ on $\Omega\tau$ for the width of AGNR =$7, 9, 12$.} 
\end{figure}
\begin{figure}
\includegraphics[width=6.5cm]{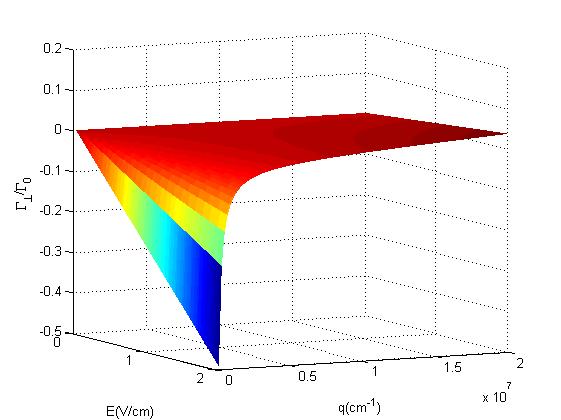}
\includegraphics[width=6.5cm]{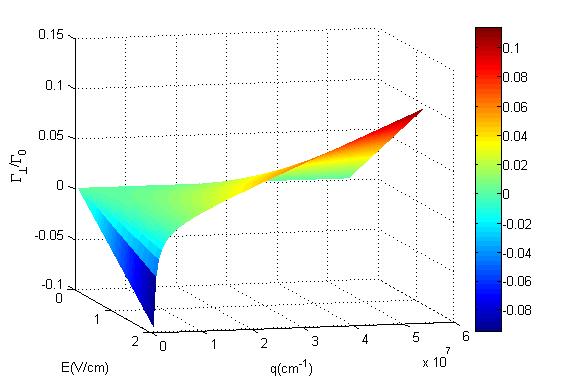}
\caption{A $3$D graph of $\Gamma/\Gamma_0$ on $E_0$ and $q$ for $p = 1$ (a) $7$-AGNR at $q = 2.0*10^6cm^{-1}$, (b) $7$-AGNR, $q = 2.5*10^6 cm^{-1}$}.
\end{figure}
\begin{figure}
\includegraphics[width=6.5cm]{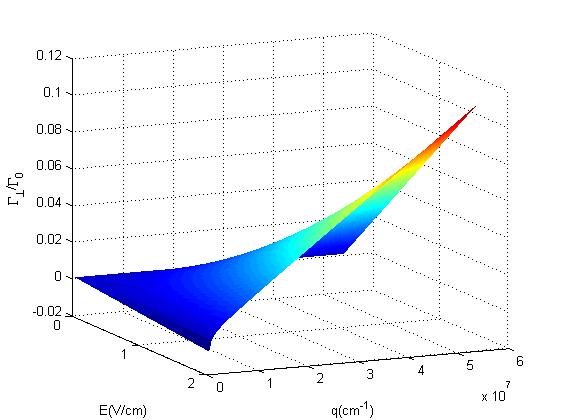}
\includegraphics[width=6.5cm]{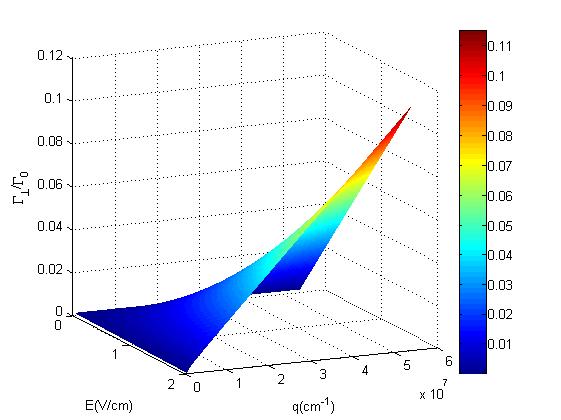}
\caption{ A $3$D graph of $\Gamma/\Gamma_0$ on $E_0$ and $q$ for  (a)$7$-AGNR at $p = 6$  (b) $8$-AGNR, at $p = 6$}.
\end{figure}
The parameters used in the numerical calculations are as follows: $\tau = 10^{-12}$s, $\omega_q =10^{10}s^{-1}$, 
$V_s = 5*10^3$ms$^{-1}$, $H = 2*10^3$ Am$^{-1}$, $q = 2.23*10$$^6cm^{-1}$.
The  Eqn.(\ref{EQ_18}) is analysed graphically for varying electric field $E_x$, acoustic wavenumber $q$,
the energy gap $E_g$ and magnetic strength ($\Omega\tau$).  Figure $1$a shows the amplification of acoustic 
waves obtained for  $7$-AGNR by varying the sub-band index $p_i = 2, 4, 6$ at specified electric fields.  
The maximum amplification was obtained at $p = 6$ but at $p = 7$, the amplification decreased. For a graph of  $\Gamma/\Gamma_0$
against $q$, Figure $1$b, showed the non-linear graph for different widths of AGNR $(7, 9, 12)$. From the graph, at   
$q < 1.5\times 10^{7}cm^{-1}$, there was an absorption. Above this value ($q > 1.5\times 10^{7}cm^{-1}$), absorption switched over to amplification and converges at higher values of $q$.     
For that of  energy gap $E_g$ against the amplification $\Gamma/\Gamma_0$ (see Fig. $2a$). The  $\Gamma/\Gamma_0$ varies 
for values of $E_g$ between $0 - 0.5$ for $7$-AGNR at $p_i= 1, 2, 3$. 
The graph of the effect of magnetic strength $(\Omega\tau)$ on  $\Gamma/\Gamma_0$ is presented in Figure $2$b. From the graph, 
$(\Omega\tau)$ increased steadily to a maximum at $0.92$ then decrease again.
In $3$D representation, the Dependence of $\Gamma/\Gamma_0$  on $q$ and $E$ are shown in Figure $3$. There is amplification 
for $7$-AGNR at $q = 2.0*10^6$ but increasing $q = 2.5*10^6$ modulates the graph. Studies of the transitions in sub-band in the AGNR by tight-binding energy dispersions 
agrees quantitatively to that of  acoustic wave amplification using Boltzmann kinetic equation. In tight-binding approximation, the electronic
structure of AGNR strongly depends on its width \cite{43}. This is verified by using $7$-AGNR and $8$-AGNR at $p = 6$ and an energy gap of $0.3eV$ (see Figure $4$).  
 The $8$-AGNR is purely absorbing but $7$-AGNR is partially amplifying. 

\section*{Conclusions}
The amplification of the acoustic wave in an external electric and magnetic field is studied using Boltzmann kinetic equation for
electron-phonon interactions in AGNR. Analytical expressions for the Amplification under different conditions are numerically analysed.
The dependence of $\Gamma/\Gamma_0$ on $E_0$ and $q$  are determined at different values of $\Omega\tau$, $p_i$ and the width where 
the maximum value of the magnetic strength occurs at $0.93$. That of $\Gamma/\Gamma_0$ against $q$ is also analysed. 
In particular, when $q$ is increased from $2.0*10^6$ to $2.5*10^6$, the amplification is modulated.

\renewcommand\refname{Bibliography}

\end{document}